# Intramembrane Cavitation as a Predictive Bio-Piezoelectric Mechanism for Ultrasonic Brain Stimulation

Michael Plaksin, Shy Shoham[*] and Eitan Kimmel[*]

*Faculty of Biomedical Engineering & Russell Berrie Nanotechnology Institute,*

*Technion – Israel Institute of Technology, Haifa 32000, Israel*

*Corresponding Authors Emails:

sshoham@bm.technion.ac.il.

eitan@bm.technion.ac.il.

**ABSTRACT.** Low-intensity ultrasonic waves can remotely and nondestructively excite central nervous system (CNS) neurons. While diverse applications for this effect are already emerging, the biophysical transduction mechanism underlying this excitation remains unclear. Recently, we suggested that ultrasound-induced intramembrane cavitation within the bilayer membrane could underlie the biomechanics of a range of observed acoustic bioeffects. In this paper, we show that, in CNS neurons, ultrasound-induced cavitation of these nanometric bilayer sonophores can induce a complex mechanoelectrical interplay leading to excitation, primarily through the effect of currents induced by membrane capacitance changes. Our model explains the basic features of CNS acoustostimulation and predicts how the experimentally observed efficacy of mouse motor cortical ultrasonic stimulation depends on stimulation parameters. These results support the hypothesis that neuronal intramembrane piezoelectricity underlies ultrasound-induced neurostimulation, and suggest that other interactions between the nervous system and pressure waves or perturbations could be explained by this new mode of biological piezoelectric transduction.

Subject Areas: Acoustics, Biological Physics, Computational Physics.

# I. INTRODUCTION

Not only is ultrasound (US) widely used for imaging [1]; its interaction with biological tissues is known to induce a wide variety of nonthermal effects ranging from hemorrhage and necrosis [2] to more delicate manipulations of cells and their membranes such as permeability enhancement [3], angiogenesis induction [4-6], and increased gene transfection [7]. In particular, both classical and recent studies have demonstrated that US can interact with the physiology of excitable tissues, inducing the generation of action potentials (APs) [8-18], suppression of nerve conduction [19-21], as well as more subtle changes in excitability [22-24]. While the suppression of nerve conduction is putatively dependent on temperature elevation [20,21], the biophysical basis of neural stimulation is not understood and has not received a rigorous, quantitative, and predictive treatment. Recently, we introduced the bilayer sonophore (BLS) model as a unifying hypothesis for the underlying mechanism of multiple bioacoustic interactions, wherein US preferentially induces intramembrane cavitation (bubble formation) in the intramembrane space between the two lipid leaflets of the cell's membrane [25]. In the BLS model, the negative pressure phase of the US wave pulls the two leaflets away while the positive pressure pushes the leaflets towards each other; dissolved gas accumulates in the hydrophobic zone, creating pockets of gas that expand and contract periodically. BLS formation is predicted to induce various alterations in cells (bioeffects) including the initiation of cellular mechano-transduction processes, the induction of membrane pore formation, and permeability changes. These bioeffects, which are naturally considered as associated with mechanical loading, were shown to systematically and predictably intensify as the US pressure amplitude increases or the frequency decreases, in softer tissues or close to free surfaces, and in the presence of microbubbles [25].

In an attempt to understand the mechanisms underlying the effect of US on excitable tissues, we analyze a neuronal BLS (NBLS) framework where the biomechanics of intramembrane cavitation is coupled to membrane bioelectrical mechanisms in a complex interplay. This mechanoelectrical coupling is shown to induce displacement currents that excite action potentials through an indirect mechanism, whose features explain the requirement for long ultrasonic stimulation pulses [12,16,18] and predict the experimentally observed efficacy of ultrasonic stimulation in mouse motor cortex [18].

## II. MODEL & EQUATIONS

The purely mechanical BLS model has been modified to account for the dynamics of membrane charge polarization, capacitance, and voltage-sensitive ion channels in a CNS neuron [*NBLS model*, Fig. 1(a)]. The NBLS model combines the modified BLS model and the Hodgkin-Huxley (H&H) model adapted for a regular spiking rat cortical pyramidal neuron [26]. Both BLS and H&H model parameters are taken "as is," without re-tuning or post-hoc adjustments, and are based on known or measured physical and biophysical quantities or ranges, wherever attainable (summarized in Table S [27]). The responses of the nanometer-scale BLS model to US are assumed to be representative of the responses of the whole cell; US waves with frequencies on the order of 1 MHz have wavelengths on the order of millimeters, orders of magnitude larger than the dimensions of CNS cortical neuron somata, so all BLS elements are subject to essentially the same acoustic effect. In the model, a circular, uniform phospholipid bilayer membrane patch is surrounded by a constraining circle of transmembrane proteins (64 nm diameter, based on average membrane interprotein distances [28]). Electrically, the bilayer membrane has a capacitance, and each ion has a Nernst equilibrium potential ($V_{Na}$, $V_K$, and $V_{Leak}$) and a time-dependent conductance, which generally depends on the product probabilities of multiple voltage-dependent gates [M and H gates for sodium channels and N and P gates for potassium channels; see Eq. (1)]. The deforming shape of the intramembrane cavity [Fig. 1(b)] is driven by the time-dependent US pressure, leaflet tension, and an attraction-repulsion equivalent pressure resulting from phospholipid molecular forces and the electrostatic attraction forces between membrane charges [top panels of Fig. 1(c) and Eq. (2)]. These dynamic deformations change the average membrane capacitance and induce a capacitive displacement current $\frac{dC_m}{dt}V_m$ [Fig. 1(c), bottom panels], which changes the membrane potential, thereby indirectly modulating the conductance of voltage-gated channels. The displacement current term is added to obtain a modified set of H&H equations for the open probability values of voltage-dependent sodium (*m* & *h*) and potassium (*n* & *p*) channels' gates:

$$\begin{aligned}
\frac{dV_m}{dt} &= -\frac{1}{C_m} \cdot \left[ V_m \cdot \frac{dC_m}{dt} + G_{Na} \cdot (V_m - V_{Na}) + G_K \cdot (V_m - V_K) + G_M \cdot (V_m - V_K) + G_{Leak} \cdot (V_m - V_{Leak}) \right] \\
\dot{n}(t) &= \alpha_n \cdot (1-n) - \beta_n \cdot n \\
\dot{m}(t) &= \alpha_m \cdot (1-m) - \beta_m \cdot m \\
\dot{h}(t) &= \alpha_h \cdot (1-h) - \beta_h \cdot h \\
\dot{p}(t) &= (p_\infty - p)/\tau_p \\
G_{Na} &= \bar{G}_{Na} \cdot m^3 \cdot h \\
G_K &= \bar{G}_K \cdot n^4 \\
G_M &= \bar{G}_M \cdot p \\
G_{Leak} &= \bar{G}_{Leak} = const
\end{aligned} \qquad (1)$$

where $G_{Na}$, $G_K$, $G_M$ and $G_{Leak}$ are the conductances of the sodium, delayed-rectiƚer potassium, slow noninactivating potassium, and leak channels, respectively, and all voltage-dependent parameters are defined in Ref. [26]. The time-dependent $C_m$ is associated with the geometrical shape of the deformed leaflets and can be determined by solving a modified dynamics force (pressure) balance equation that is based on the

Rayleigh-Plesset (RP) equation for bubble dynamics [25,29]. The modified BLS equation,

$$\frac{d^2Z}{dt^2} + \frac{3}{2R(Z)}\left(\frac{dZ}{dt}\right)^2 = \qquad (2)$$

$$\frac{1}{\rho_l |R(Z)|}\left[P_{in} + P_M + P_{ec} - P_0 + P_A \sin(\omega t) - P_S(Z) - \frac{4}{|R(Z)|}\cdot\frac{dZ}{dt}\left(\frac{3\delta_0\mu_s}{|R(Z)|} + \mu_l\right)\right],$$

is the original dynamics equation [25] with a new electric equivalent pressure term,

$$P_{ec} = -\frac{a^2}{Z^2 + a^2}\frac{(C_m \cdot V_m)^2}{2\varepsilon_0\varepsilon_r} \qquad (3)$$

as well as slightly changed molecular forces between the phospholipid molecules in the opposite leaflets,

$$f(r) = A_r\left[\left(\frac{\Delta^*}{2z(r)+\Delta}\right)^x - \left(\frac{\Delta^*}{2z(r)+\Delta}\right)^y\right] \qquad (4)$$

that are captured by an equivalent "molecular pressure,"

$$P_M = \frac{2}{Z^2 + a^2}\int_0^a f(r)r\,dr \qquad (5)$$

where $a$, $Z$, $z(r)$ and $R(Z)$ are defined in Fig. 1; $\rho_l$ is the membrane surrounding water-like medium density; $\varepsilon_0$ and $\varepsilon_r$ are the dielectric and relative dielectric constants, respectively, of vacuum and of the transmission medium between the two leaflets ($\varepsilon_r$ chosen here to be equal to 1), and $\Delta$, $\Delta^*$ are the initial gaps between the leaflets if there are or there are no charges on the membrane, respectively. Equation (3) accounts for the effective pressure due to the attraction forces between the electric ion charges on the membrane leaflets. The other pressure terms in Eq. (2) account for the surface tension [25] in the leaflets ($P_S$) and for the intramembrane gas (air; $N_2+O_2$) pressure ($P_{in}$). The rest pressure is $P_0$ and the driving pressure is the US pressure $P_A \sin(\omega t)$. The last pressure term on the rhs of Eq. (2) accounts for the viscous loss [25].

The value of $P_{in}$ is determined by the following gas balance equation:

$$\frac{dn_a}{dt} = \frac{2 \cdot \pi \left(a^2 + Z^2\right) D_a}{\xi} \left[ C_a - \frac{P_{in}}{k_a} \right],$$

$$n_a \big|_{t=0} = \frac{P_0 \pi a^2 \Delta}{R_g Tem}, \qquad (6)$$

$$P_{in} = \frac{n_a R_g Tem}{V_a}$$

Equation (6) presents gas transport between two compartments: i) the dissolved gas compartment, with a uniform gas concentration ($C_a$) in the gas-saturated water, where $D_a$ is the diffusion coefficient of the air in the water; and ii) the intra-membrane gaseous compartment, where $k_a$ is the Henry coefficient, $n_a$ is the mole content of ideal gas, $R_g$ is the gas constant, $Tem$ is the temperature, and $V_a$ is the intramembrane cavity volume, which is expressed by

$$V_a = \pi a^2 \Delta \left[ 1 + \frac{Z}{3\Delta} \left( \frac{Z^2}{a^2} + 3 \right) \right] \qquad (7)$$

The gas transport takes place across a boundary layer with thickness $\xi$ near the leaflet. For simplicity, we assume that $\xi = 0.5$ nm and as a result the gas reaches almost immediate equilibrium oscillations between the dissolved and the gaseous compartments. Such an assumption saves much time, avoiding the complex calculations of the air concentration field in the surrounding medium, and is justified by the relative short times that the intramembrane space requires to reach stable oscillations [25]. We also assume that [$P_{in}(t=0) = P_0 = 10^5$ $Pa$] and that the initial gap between the leaflets is $\Delta = 1.26$ nm.

As the two leaflets of the BLS separate, deform, and curve periodically with the US pressure, the membrane capacitance in Eqs. (1) and (3) is approximately derived by

$$C_m(Z) = \frac{1}{\pi a^2} \int_0^{2\pi}\int_0^a \left(\frac{C_{m_0} \cdot \Delta}{2z(r)+\Delta}\right) r\, dr\, d\theta = \frac{C_{m_0} \cdot \Delta}{a^2}\left(Z + \frac{(a^2 - Z^2 - Z\Delta)}{2Z}\ln\left[\frac{2Z+\Delta}{\Delta}\right]\right) \quad (8)$$

which is based on a parallel-plate capacitor expression per unit area and where the $C_{m_0}$ is cell membrane capacity at rest.

## III. RESULTS

### A. Fundamental response to ultrasound and AP generation

We first studied the NBLS model's fundamental response to US stimulation. When continuous-wave (CW) US stimulates the model neuron, the intramembrane space inflates and deflates at the US frequency [25], and the NBLS membrane potential oscillates strongly between -280 and -60 mV (Fig. 2, 0.35 MHz CW pulse, acoustic pressure amplitude 100 kPa, intensity 320 mW/cm$^2$ in a propagating wave). These membrane potential oscillations are driven by the acoustic-frequency capacitive currents, while the much-faster sign-changing capacitive current oscillations seen during the leaflets' closing [Fig. 1(c)] do not accumulate and lead only to negligible variation in the membrane potential [Fig. 2(a), inset]. When the US stimulation stops after 30 ms, a single AP is generated after a short latency [Fig. 2(a)]. For increased US pulse duration [Fig. 2(b), 40 ms], the US stimulation leads to the generation of several APs *during* the stimulus (but starting at a later time). What is the detailed biophysical basis underlying AP generation in this model? Before the onset of US, the resting membrane potential is -72 mV, for which the voltage-dependent M, N, and P gates are almost closed while the H gate is wide open. The US-induced oscillations are hyperpolarizing and further decrease the membrane potential. [The transient periods of higher membrane potential are too short for the channels' response-time constants; Fig. 2(a), bottom panels.] At this stage, all voltage-dependent ion channels are closed; however, non-voltage-dependent ion channels remain open, and fluctuating leak currents enter and exit the cell, causing a net elevation in the membrane's charge. When the US stimulation stops (after 30 ms), the membrane capacitance returns to its reference value, and the membrane potential is determined by the accumulated charge. The M gates are now the first to respond to the membrane potential change-if it exceeds roughly -50 mV, a single AP is rapidly generated [Fig. 2(a)]. When the pulse duration is increased [40ms, Fig. 2(b)], US stimulation leads to the generation of a few APs during the stimulus, primarily due to the gradual shift of the membrane potential oscillation range towards more depolarized values (e.g., from -280 to -60 mV at US onset and -160 to -35 mV, 33 ms later), gradually increasing the conductances of both sodium and potassium channels to relatively high values [Fig. 2(b), insets]. This detailed mechanism is reminiscent of but very different from anodal-break neuronal excitation [30]; US-induced excitation is not related to the inactivation state of the H gates, but rather it is driven by the inability of channels to respond at the rates at which US-induced oscillations occur. (In fact, anodal break does not occur in this particular neuron model [26].)

### B. Dependence on ultrasound parameters

Next, we examined the influence of US frequency and duration on the threshold CW US intensity (and energy) required to generate an action potential and the number of APs generated. Intensity activation thresholds were found to monotonically (but weakly) increase with US frequency [Fig. 3(a)] and decrease with pulse duration at a fixed stimulation frequency [0.35 MHz, Fig. 3(b)], while threshold activation *energy* per unit area (intensity x duration) reaches a minimum at intermediate durations [~50 ms, Fig. 3(c)]. This frequency dependence is associated with increased inertia of the surrounding media against motion at higher frequencies, while the activation energy minimum is due to a tradeoff between charge accumulation and charge leakage as well as the relatively high molecular forces at low intensities. Interestingly, excitation by direct injection of a short 1ms current pulse to the model neuron was found to require approximately 5 orders of magnitude less intensity (approximately 1.3 µW/cm$^2$) and about 6 orders of magnitude less energy (approximately 1.3 nJ/cm$^2$) to elicit an action potential, highlighting the low energetic efficiency of the mechanical excitation process.

Above threshold, the number of APs generated increases monotonically with both intensity and duration [Fig. 3(d)]; as US intensity rises, the firing rate increases [Fig. 3(e)], while the latency to the first action potential decreases [Fig. 3(f)]. Both the rate and the latency are essentially frequency independent at frequencies below 1 MHz [Fig. 3(e) and 3(f)]. This behavior can be explained by the relatively minor changes in threshold intensity for a single AP generation over the range of 0.2-4 MHz [Fig. 3(a)].

### C. Biophysical model predicts *in vivo* results

Finally, to validate the NBLS model, we compared its predictions to the results of a recent *in-vivo* study [18] where a wide range of US parameters were used to stimulate mouse primary motor cortex while resulting front limb muscle EMG signals were measured. To compare model predictions with experimental measurements we used Buckingham-Pi dimensional analysis [31] to relate NBLSø measures ó the number of AP spikes ($N$), the response latency ($L$), and the overall duration of APs ($D$), to the experimental success rate ($R_{sr}$). Two dimensionless variables, $N$ and the response

"effectiveness" $D/(D+L)$, are associated with success rate in our model, but the latter quantity also appears to be solely a function of $N$ (Fig. S1 [27]). Therefore, $R_{sr}$ depends only on $N$, a dependence that is well approximated by a sigmoidal-shaped logistic function (see the Appendix and Fig. S2 [27]), and $R_{sr}(N)$ can be calibrated using this functional fit.

The calibrated NBLS model output was used to predict success rates for varying US frequencies and intensities under the conditions used by King *et al.* [18] : Model predictions clearly show a high degree of qualitative agreement with their results [Fig. 4(a); stimulation pulses have 40,000 cycles and different frequencies]. Similar qualitative agreement is seen for 0.5 MHz US stimuli with varying pulse durations and pressure amplitudes (shown relative to the maximal used amplitude of 725 kPa or 16.8 W/cm$^2$). For relatively short pulse durations and low US pressure amplitudes, a relatively low success rate is predicted [blue area in Fig. 4(b)]. When the pulse duration or the amplitude increases, the success rate increases gradually with power and duration until it reaches a plateau of 100 % success rates.

A final simulation study examined excitation by 0.5 MHz US *pulse trains* for different intensities and duty cycle values. (Pulse-mode US is commonly used in applications in which it is desirable to avoid heating the tissue, including neural stimulation.) We have found that the AP excitation mechanism for pulsed excitation is generally the same as for the CW mode (Fig. 2). Success probability for CW excitation is predictably higher than for 30 % duty-cycle pulsed excitation, a trend which is consistent with the experimental results [18]. Interestingly, however, a certain intermediate duty cycle value (about 70 %) is predicted to have a maximal success rate, higher than for CW excitation [Fig. 4(c)].

## IV. DISCUSSION

We studied a new biomechanical-biophysical mode of interaction between US and neurons with the goal of investigating the non-destructive manipulation of excitable tissues by US [11]. Such an understanding can guide the development of future therapeutic applications of the only technology currently capable of *targeted non-invasive brain stimulation*. Exposing the NBLS element to simulated US results in rapid oscillatory hyperpolarizing currents and can lead to AP generation through a charge accumulation mechanism that results from the imbalance of ionic currents. This indirect, energetically inefficient excitation mechanism explains the characteristically very long pulses required for acoustostimulation - tens to hundreds of milliseconds [12,16,18], compared to submillisecond pulses typically used for electrical stimulation or even for photothermal stimulation [32,33] (recently shown [34] to analogously be induced by *membrane capacitance changes* leading to depolarizing currents). This model also explains qualitatively how *off* responses can be (commonly) elicited after stimulus termination [17] [Fig. 2(a)]. Detailed model predictions were found to qualitatively agree with the results of recent *in-vivo* experiments in mice [18] (Figs. 4 & Fig. S3 [27] - The first study in which the effectiveness of a range of stimulation parameters was systematically examined). The agreement is potentially *much closer* (Fig. S4 [27]) when taking into consideration that their pressure calibrations with an *ex-vivo* mouse skull cap have likely underestimated the actual pressures generated by standing waves inside the skull. This result lends some support to our underlying assumptions, simplifications, and the natural choice of (unadjusted) parameters: In particular, it putatively supports using the response of a nanometric NBLS as predictive of the compound average behavior of the whole CNS neuron and of a whole neuronal population exposed to US (a realistic assumption given the millimeter-scale beam dimensions and wavelengths). While the assumption of structured intra-membrane cavities appears perhaps unrealistically simplistic, observed protein distributions in real cells' membranes are in fact clustered and somewhat reminiscent of these idealized circular distributions (see, e.g., Fig. 1 in Ref. [35], where protein-free patches of 50-100nm diameters are evident). The current NBLS model captures only excitatory regular-spiking pyramidal cortical neurons. These are not only the most common neocortical neuron type [26], but they also account for the efferent output onto the motor periphery [36]. It is also worth noting that intramembrane cavitation is not the only possible underlying mechanism for US-induced excitation; for example, the flexoelectricity model is a popular choice for explaining biological mechanoelectrical conversion [37] and is also based on varying the curvature of the bilayer membrane. In Petrov's model, the two membrane leaflets deform together as one sheet with roughly the same radius of curvature, while in our model, the distance between the leaflets varies as they expand (and collapse) away from each other [Fig. 1(b)]. Our model is thus much more sensitive to acoustic pressure (acting *directly* to deform the BLS [25]) while deformations in the membrane's curvature from the negligible acoustic radiation pressures induced across the cell's membrane require unrealistic intensities for flexoelectric depolarizations of several millivolts. Another option we have considered is a direct influence of membrane tension on voltage-gated channels; however, in simulations incorporating a tension-dependent rate constant (inferred from Ref. [38]), US-induced effects were only minor and were orders-of-magnitude away from leading to excitation (data not shown). Naturally, future studies should also consider the role of acoustic interactions with other types of cortical cell (and potentially also cells in the thalamus) in the observed excitation, as well as effects on specific subcellular mechanosensitive elements including ion channels like MmPiezo1 and MmPiezo2 [39], presynaptic calcium channels [17,40], or postsynaptic NMDA and AMPA glutamate receptors [41].

The NBLS framework not only agrees with but also sheds new light on experimental results. While the dependence of success rate on frequency [Fig. 4(a)] can be interpreted as strong [18], the model suggests that it is primarily a result of different pulse durations used experimentally, while both number of APs and success rate are almost independent of the frequency [Figs. 3(e) and 3(f)]. The model also predicts an optimal duty cycle for pulsed excitation [Fig. 4(c)], reflecting an optimal tradeoff between charge accumulation and M gates ability to respond (preferentially during stimulus breaks). This result also highlights the potential of this model-based approach in the design of optimal acoustic excitation waveforms and strategies.

Finally, we speculate that this new mode of membrane-based US-induced piezoelectric transduction should exist in all polarized bilayer membranes, and it could be interesting to explore its properties, short- and long-term effects, and applications both in and out of the nervous system. Indeed, the electromechanical "capacitive displacement current" term $\frac{dC_m}{dt}V_m$ (which does not appear in the H&H equations) does appear to play a role in other contexts as well. For example, Heimburg and Jackson [42] propose an alternative electromechanical basis for action potential propagation that uses these capacitive currents and which is consistent with the action-potential-induced transient swelling observed by Iwasa, Tasaki, and Gibbons [43]. Similar interactions to the ones we have studied may underlie and/or shed new light on other types of biological mechanoelectric coupling, from the sensation of sound vibration in the auditory system (where sub-Pascal-scale threshold-pressure amplitudes [44] could rely on hypersensitized NBLS-type mechanisms) and up to the impact of mega-Pascal pressure shocks that cause concussions and traumatic brain injury [45-47].

**ACKNOWLEDGMENTS**

We thank Yoav Henis, Roman Shusterman, Joshua Goldberg, Omer Naor, Steve Krupa, Amit Livneh, and two anonymous reviewers for their comments on the manuscript. This work was supported by the Russell Berrie Nanotechnology Institute, Johnson & Johnson Grant No. 1010051, and European Research Council Starting Grant No. 211055.

**APPENDIX: MODEL SIMULATION AND VALIDATION DETAILS**

The model's set of equations was numerically solved in MATLAB (using the function ODE113). The time distance between the calculated points is set to $0.025/f$ µs (where $f$ is the US frequency in MHz). The modified BLS model [Eqs. (2), (3), (4), (5), (6) and (7)] was solved simultaneously with the H&H modified equations [Eqs. (1) and (8)] by updating the charge $C_m V_m$ in equation (3) every 500 µs. In each simulation run, the BLS model was solved as long as $Z(t)$ and $n_a(t)$ kept evolving in time. When both functions reached a stabilized periodic solution, we used Fourier series to incorporate it into Eq. (1).

For comparison between the theory and the experiments we introduced a logistic function,

$$R_{sr} = \frac{100}{1+e^{\beta_0 + \beta_1 N}} \quad (A1)$$

where $\beta_0$ and $\beta_1$ can be estimated from the *success rate ($R_{sr}$)* versus intensity (Table I in Ref. [18]), and the *number of AP spikes (N)* versus intensity predicted by the NBLS model at US frequency of 0.5 MHz. Two data points were used ($R_{sr}$ in percentage, $N$) (28.7, 34) and (52.9, 45). The calibration curve can be seen in Fig. S2 [27].

The intensity (*I*) in a propagating US CW [48] was calculated by:

$$I = \frac{P_A^2}{2\rho_l c} \qquad (A2)$$

where $P_A$ is the pressure amplitude, $\rho_l$ is the surrounding medium density, and *c* is the speed of sound in the medium. The spatial-peak-pulse average intensities in the pulsed-mode simulations were calculated using Eq. (A2) as well.

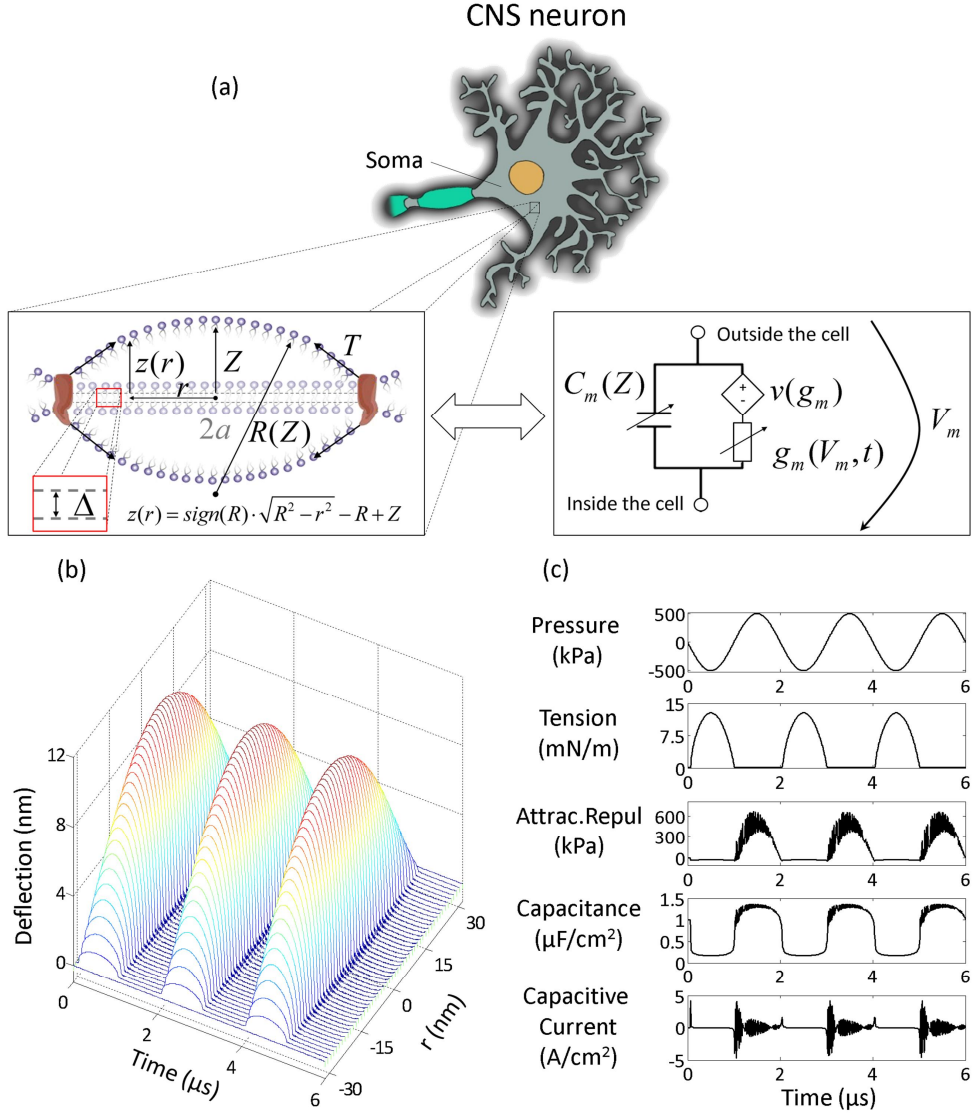

FIG. 1. Neuronal bilayer sonophore model. (a) Biomechanical and bioelectrical structure of NBLS model of small membrane patch in a CNS pyramidal neuron. A round patch of the bilayer leaflets (radius *a*, initial gap *Δ*) dynamically deforms into a dome shape with a maximal deflection *Z*, radius of curvature *R(Z)* and tension *T*. The membrane equivalent circuit has a potential ($V_m$), time-varying capacitance ($C_m$), and Hodgkin-Huxley type ionic conductances ($g_m$) and sources [$v(g_m)$]. (b,c) Mechano-electrical dynamics of first three cycles of the model membrane exposed to US (pressure amplitude 500 kPa and frequency 0.5 MHz): (b) Local deflection at each radial coordinate [$z(r)$]. (c) Acoustic pressure (kPa), tension (mN/m), combined attraction/repulsion force per area between the leaflets (sum of molecular and electrostatic forces $P_M+P_{ec}$, kPa), membrane capacitance ( F/cm$^2$), and capacitive displacement current $\frac{dC_m}{dt}V_m$ (A/cm$^2$).

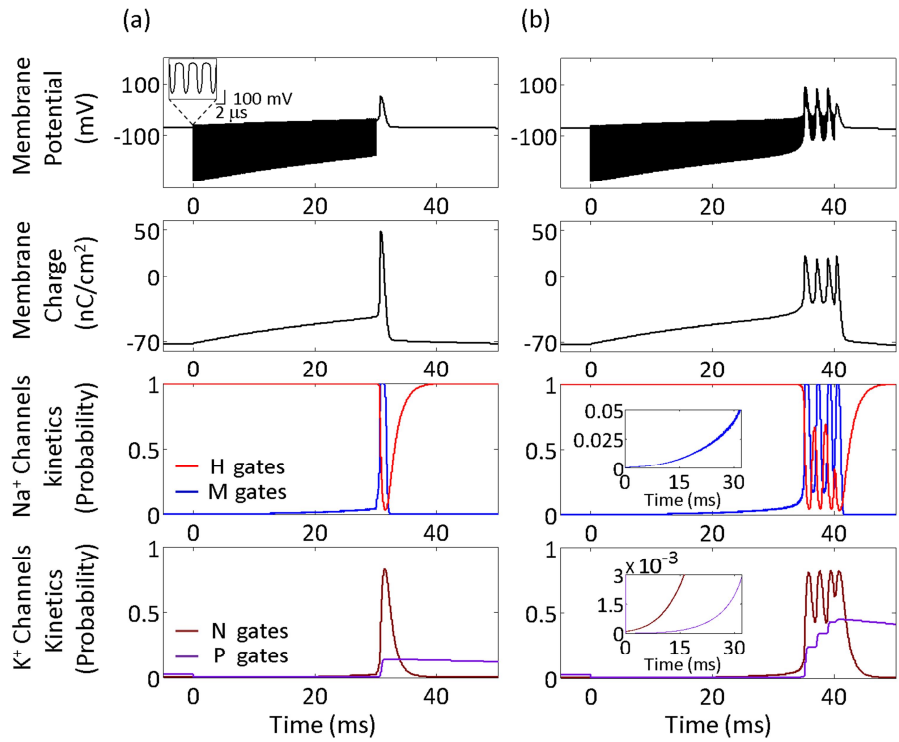

FIG. 2. The effect of continuous US stimuli (frequency 0.35 MHz, intensity 320 mW/cm$^2$) on membrane potential & charge (top panels), and on sodium and potassium channels kinetics (bottom). (a) For stimulus duration of 30 ms, a single AP is generated immediately after the stimulus ends. Inset shows membrane potential oscillations. (b) For a stimulus duration of 40 ms, four APs are generated during the stimulus. Insets: magnified view of open probabilities for the M gate (Na$^+$ channel kinetics) and the N and P gates (K$^+$ channels).

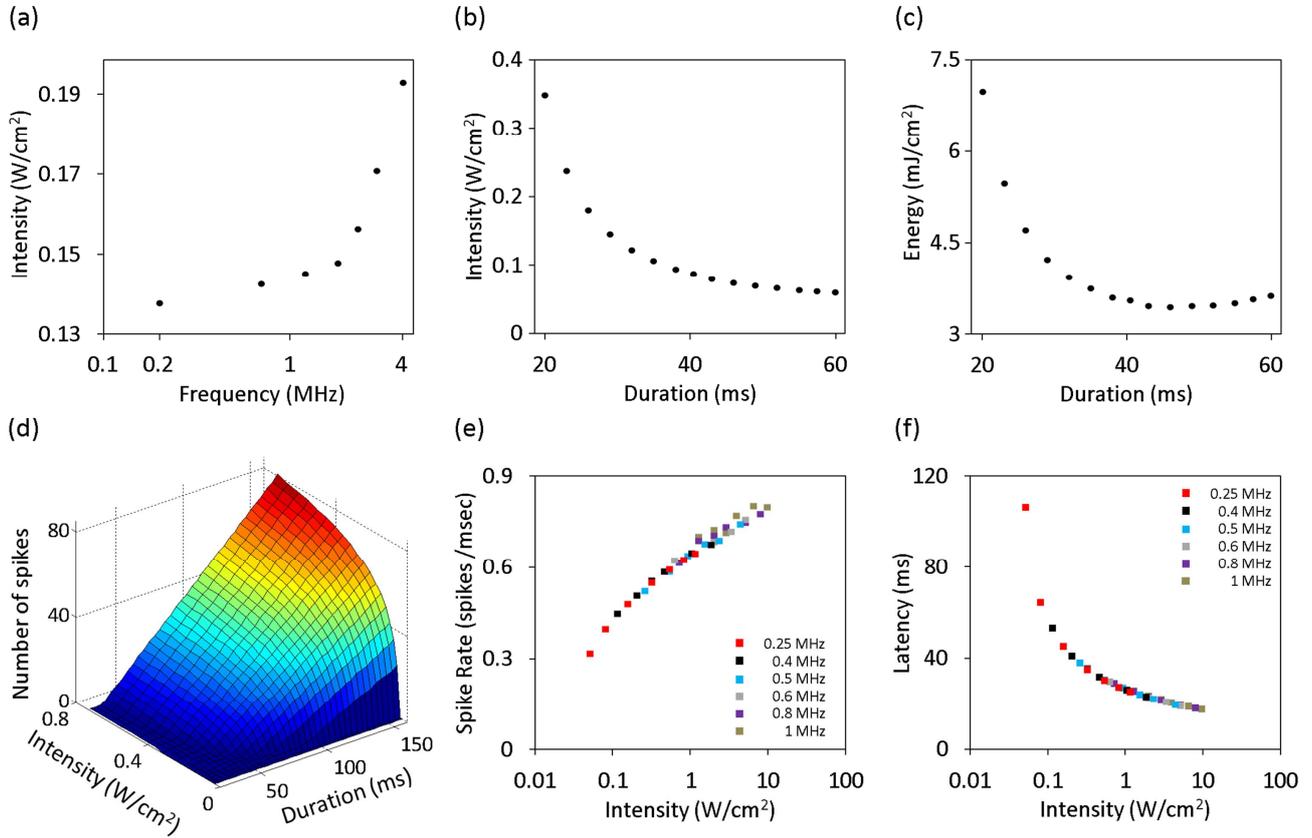

FIG. 3. Relationship between US stimulus parameters and APs generation in NBLS model. (a) Threshold intensity versus frequency required to generate a single AP (duration: 30 ms). (b,c) Threshold intensity and energy per unit membrane area versus stimulus duration required to generate a single AP (frequency: 0.35 MHz). (d) The number of APs induced by US stimuli as a function of US intensity and stimulus duration (frequency: 0.35 MHz). (e) Spike rate versus intensity for US frequencies (in MHz) of 0.25 (red squares), 0.4 (black squares), 0.5 (blue squares), 0.6 (gray squares), 0.8 (purple squares) and 1.0 (brown squares). (f) Latency before the appearance of the first AP during the US stimulation versus US intensity. Symbols and colors as in *(e)*.

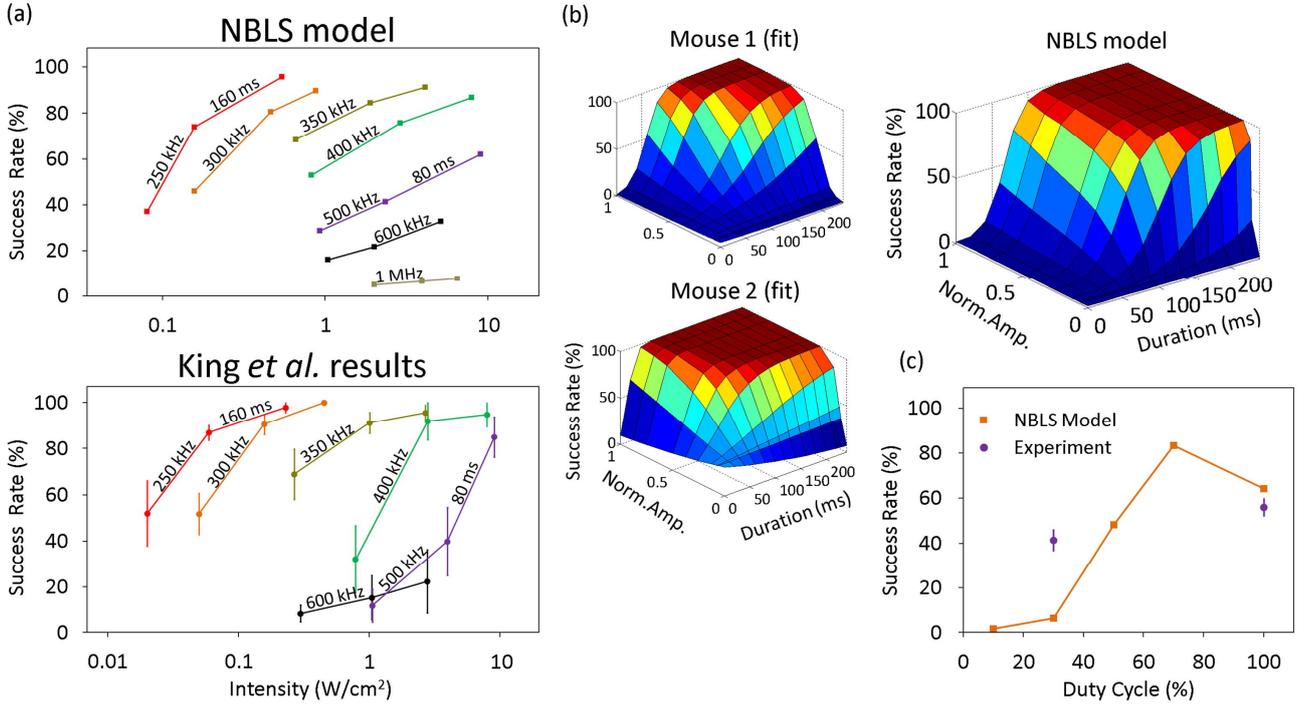

FIG. 4. Comparison of NBLS model predictions and *in-vivo* brain stimulation measurements [18]. (a) Success rates for eliciting motor responses versus US intensity at different frequencies (continuous stimulation for 40000 cycles). NBLS predictions (top panel) and the experimental results from Ref. [18] (bottom panel) are shown for US frequencies in the range 250 to 600 kHz (an additional 1 MHz simulation is also shown). Note that stimulus duration changes with the frequency (representative values shown). (b) Model-based predictions of success rate as a function of US pressure amplitude and stimulus duration (pressure amplitudes were normalized by maximal value used, 725 kPa; frequency: 0.5 MHz). For comparison, *in-vivo* measurements in two different mice [18] are shown in the upper part for the same US stimulation parameters. (c) Success rate as a function of duty cycle for simulation of NBLS model with pulsed-mode US (frequency: 0.5 MHz, Pulse repetition frequency: 1.5 kHz, stimulus duration: 80 ms, intensity: 10 W/cm$^2$ spatial peak pulse average). Two experimental data points from Ref. [18] are shown for duty cycle values 30 % and 100 % (stars, other US parameters were similar to the simulated values). Error bars represent the standard error of the mean.

# Intramembrane Cavitation as a Predictive Bio-Piezoelectric Mechanism for Ultrasonic Brain Stimulation

*Supplemental Material*

Michael Plaksin, Shy Shoham* and Eitan Kimmel*

*Faculty of Biomedical Engineering & Russell Berrie Nanotechnology Institute,*

*Technion – Israel Institute of Technology, Haifa 32000, Israel*

*Corresponding Authors Emails:

sshoham@bm.technion.ac.il.

eitan@bm.technion.ac.il.

Table SI. The parameters for the simulation runs of the NBLS model.

| Parameter | Symbol | Unit | Value | Source |
|---|---|---|---|---|
| Thickness of the leaflet. | $\delta_0$ | nm | 2 | [S1] |
| Initial gap between the two leaflets if there is no charge on the membrane. | $\Delta^*$ | nm | 1.4 | [S2] |
| Initial gap between the two leaflets if there is a charge on the membrane. | $\Delta$ | nm | 1.26 | Calculated from equilibrium state¶ |
| Attraction/repulsion pressure coefficient. | $A_r$ | Pa | $10^5$ | [S2] |
| Exponent in the repulsion term. | $x$ | - | 5 | [S2] |
| Exponent in the attraction term. | $y$ | - | 3.3 | [S2] |
| Dynamic viscosity of the leaflets. | $\mu_s$ | Pa·s | 0.035 | Educated guess |
| Dynamic viscosity of the surrounding medium. | $\mu_l$ | Pa·s | $0.7 \cdot 10^{-3}$ | [S3] |
| Diffusion coefficient of air in the surrounding medium. | $D_a$ | m²·s⁻¹ | $3 \cdot 10^{-9}$ | [S4] |

| Parameter | Symbol | Unit | Value | Source |
|---|---|---|---|---|
| Density of the surrounding medium. | $\rho_l$ | kg·m$^{-3}$ | 1028 | [S5] |
| Speed of sound in the surrounding medium. | $c$ | m·s$^{-1}$ | 1515 | |
| Initial air molar concentration in the surrounding medium (O$_2$+N$_2$). | $C_a$ | mol·m$^{-3}$ | 0.62 | [S3,S6,S7] |
| Henry's constant for dissolved air in the surrounding medium. | $k_a$ | Pa·m$^3$·mol$^{-1}$ | 1.63·10$^5$ | |
| Static pressure in the surrounding medium. | $P_0$ | Pa | 10$^5$ | |
| The radius of the leaflets' boundary. | $a$ | nm | 32 | Selection based on [S8] |
| The boundary layer length between the surrounding medium and the leaflets. | $\xi$ | nm | 0.5 | Educated guess |
| The areal modulus of the bilayer membrane. | $k_s$ | N·m$^{-1}$ | 0.24 | Selection based on [S9] |
| Surrounding medium temperature. | $Tem$ | K | 309.15 | [S10] |
| The maximal conductance of the Na$^+$ channels. | $\bar{G}_{Na}$ | mS·cm$^{-2}$ | 56 | |
| The maximal conductance of the delayed-rectifier K$^+$ channels. | $\bar{G}_K$ | mS·cm$^{-2}$ | 6 | |
| The maximal conductance of the slow non-inactivating K$^+$ channels. | $\bar{G}_M$ | mS·cm$^{-2}$ | 0.075 | |
| The maximal conductance of the non-voltage-dependent, non-specific ions channels. | $\bar{G}_{Leak}$ | mS·cm$^{-2}$ | 0.0205 | |
| The Nernst potential of the Na$^+$ | $V_{Na}$ | mV | 50 | |
| The Nernst potential of the K$^+$ | $V_K$ | mV | -90 | |
| The Nernst potential of the non-voltage-dependent, non-specific ion channels. | $V_{Leak}$ | mV | -70.3 | |
| Spike threshold adjustment parameter. | $V_T$ | mV | -56.2 | |
| The decay time constant for adaptation at slow non-inactivating K$^+$ channels. | $\tau_{max}$ | ms | 608 | |
| The cell membrane capacity at rest. | $C_{m_0}$ | μF·cm$^{-2}$ | 1 | |
| The rest potential of the cell membrane. | $V_{m_0}$ | mV | -71.9 | |

| Parameter | Symbol | Unit | Value | Source |
|---|---|---|---|---|
| Relative permittivity of the intramembrane cavity. | $\varepsilon_r$ | - | 1 | Educated guess |
| Logistic function parameter. | $\beta_0$ | - | 4.08 | Calculated from [S5] |
| Logistic function parameter. | $\beta_1$ | - | -0.093 | |

¶ $\Delta$ is considered to be an *effective* initial distance parametrizing the inter-leaflet charge separation in the equivalent plate capacitor. Two different estimates led to essentially the same result: (1) from the membrane's capacitance ~1μF/cm² for a parallel plate capacitor: $C = \dfrac{\varepsilon_r \varepsilon_0}{d}$, one obtains that $d_{eff}$ is somewhere between 1.77nm (for $\varepsilon_r \approx 2$, of a pure lipid-bilayer [S11]) and 0.89nm (for the air-filled membrane gap where $\varepsilon_r \approx 1$). Thus, an intermediate value of ~1.3nm is a reasonable estimate for the effective gap in a mixed lipid-air membrane capacitor. (2) the membrane leaflets' equilibrium state occurs at $\Delta$ =1.26nm [equating the sum of Eqs. (3)+(5) to zero when Z=0].

Ä The patch's radius *a* was derived from the *average distance between neighboring proteins* (53 nm) in native oocytes [S8]. In a rectangular grid geometry, such an average inter-protein distance corresponds to a radius of 31±1 nm (half of the unit cell's diagonal ± a derived error estimate, of which we chose the maximal, least stiff value 32nm).

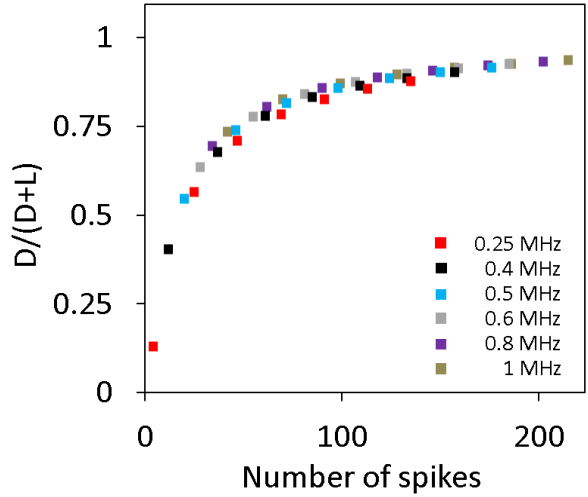

FIG. S1. The NBLS response "effectiveness" curve. Simulation results for $D/(L+D)$ as a function of the number of spikes for US frequencies of 0.25, 0.4, 0.5, 0.6, 0.8 and 1 MHz. One can observe the weak dependence on US frequency.

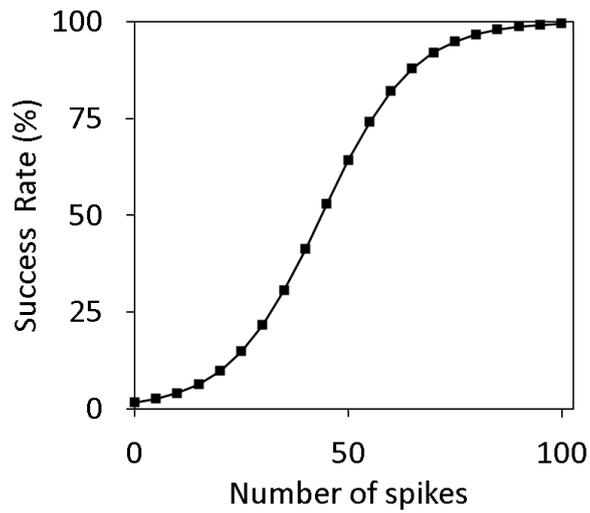

FIG. S2. Calibration curve between the NBLS model and the experimental results [S5]. Calibration curve of $R_{sr}$ vs. $N$, where $R_{sr}$ is the success rate and $N$ is the number of spikes.

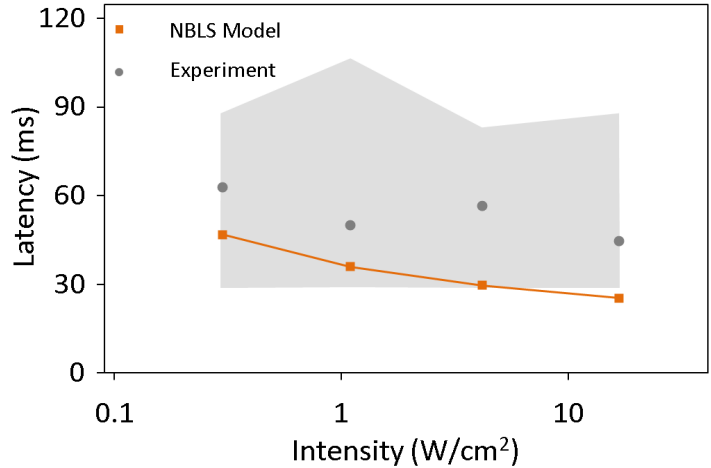

FIG. S3. Comparison of experimental vs. model-predicted latencies of US-evoked EMGs for different US intensities. The figure compares data derived from King *et al.*'s latency histograms (gray circles indicate distribution mode and the gray zone is the range between the half-maxima: data from Fig. 6 of Ref. [S5]) and the latencies derived from the NBLS model (orange squares), adjusted by a 10ms signal propagation delay between the motor cortex and the forelimbs' muscles [S12]. Stimulus parameters: CW, duration - 80 ms and US frequency - 0.5 MHz.

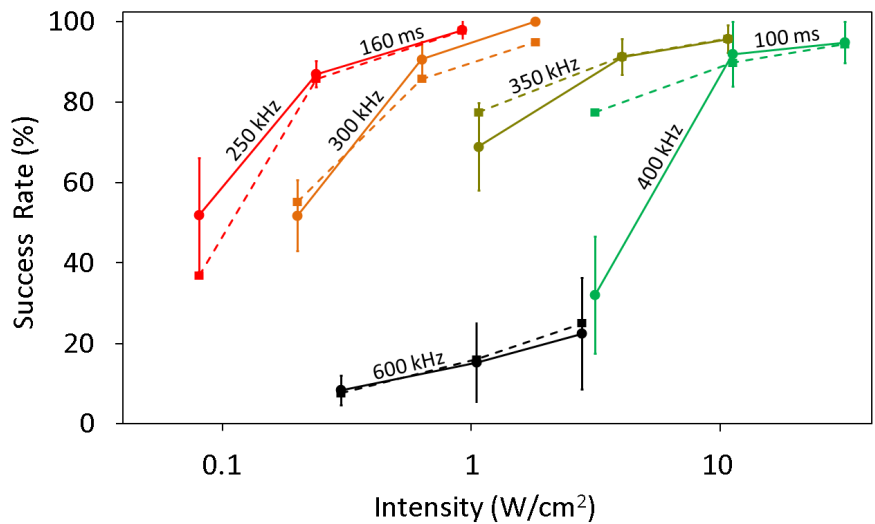

FIG. S4. Comparison between NBLS model (full squares and dashed lines) and experimental results heuristically adjusted to account for the generation of standing waves in the skull (full circles and solid lines). To account for standing waves expected at low US frequencies (e.g., 0.25-0.4 MHz), we doubled the pressure amplitudes used by King *et al.* [S5] only at those frequencies (intensities multiplied by a factor of 4). For the comparison, we used the logistic function calibration described in the main text (appendix). Error bars represent the standard error of the mean.

**REFERENCES FOR SUPPLEMENTAL MATERIAL**